# Atomic-Layer-Deposited $Al_2O_3$ on $Bi_2Te_3$ for Topological Insulator Field-Effect Transistors

Han Liu and P.D. Ye [a]

*School of Electrical and Computer Engineering and Birck Nanotechnology Center, Purdue University, West Lafayette, IN 47907*

(August 4, 2011)

We report dual-gate modulation of topological insulator field-effect transistors (TI FETs) made on $Bi_2Te_3$ thin flakes with integration of atomic-layer-deposited (ALD) $Al_2O_3$ high-k dielectric. Atomic force microscopy study shows that ALD $Al_2O_3$ is uniformly grown on this layer-structured channel material. Electrical characterization reveals that the right selection of ALD precursors and the related surface chemistry play a critical role in device performance of $Bi_2Te_3$ based TI FETs. We realize both top-gate and bottom-gate control on these devices, and the highest modulation rate of 76.1% is achieved by using simultaneous dual gate control.

Three dimensional topological insulator (TI) materials, such as $Bi_2Te_3$, $Bi_2Se_3$, and $Sb_2Te_3$, have recently attracted much attention due to their unique physical properties.[1-5] These structures are layered like graphene but appear to behave like insulators, having a band gap in the material bulk. However, these materials show metallic behavior on their surfaces. The surface states of TIs consist of an odd number of massless Dirac cones, around which the linear dispersion of the electron spectrum is such that the carriers reach extremely high surface carrier mobilities up to 9000-10000 $cm^2$/Vs.[6] Moreover, these surfaces, protected by time reversal symmetry,[7] result in a non-scattering carrier transport regime, making TIs rather promising for future nanoelectronics applications with ultralow power dissipation.

In order to realize practical TI field-effect transistors (FETs), it is important to study the integration of gate dielectrics on these materials so that the channel current can be controlled by the top-gate. In some early studies of $Bi_2Te_3$ based TI devices, where a heavily doped silicon back gate was used, no modulation was observed within a back-gate voltage sweep from -50V to 50V at room temperature.[8] The highest modulation obtained is around 20% measured at less than 10 K.[9] For $Bi_2Se_3$, which has a larger bandgap of 0.3 eV than 0.14 eV of $Bi_2Te_3$, minuscule gate modulation was reported using top-gate control at room temperature,[10] while much larger modulation was achieved by the back-gate sweeps.[11] Although these are inspiring results to observe the field effect by using a global back-gate, this cannot satisfy the requirement for real device applications, which requires individual device top-gate control and room temperature operation. Therefore, the realization of highly-efficient top-gate modulation for $Bi_2Te_3$ and other TI materials at room temperature is urgently needed.

In this letter, we demonstrate $Al_2O_3$ growth by atomic-layer deposition (ALD) with two different precursors on $Bi_2Te_3$ top surfaces. The uniformity of the surface morphology after the ALD $Al_2O_3$ deposition is also studied. Electrical characterization has shown a strong modulation of $Bi_2Te_3$ based TI FETs with both top-gate and back-gate controls. $Bi_2Te_3$ thin flakes were peeled off from bulk ingots by standard 3M scotch tape techniques[12] and were then transferred to a highly doped silicon wafer with 300 nm thick $SiO_2$. The thickness of these ultrathin flakes is less than 50 nm. Ten nanometer $Al_2O_3$ high-k dielectric layers were deposited by an ASM F-120 ALD system at 200$^o$C by using tri-methyl-aluminum (TMA) and $H_2O$ or TMA and $O_3$ as precursors. The pulse time is 0.8s TMA and 1.2s $H_2O$, or 1s TMA and 1s $O_3$, and the purge time is 6s $N_2$ for each precursor. Source/drain regions were defined by optical lithography. $Al_2O_3$ was etched away using buffered oxide etch (BOE) at these source/drain regions, followed by e-beam evaporation of a 20nm/40nm Cr/Au metal and lift-off process for ohmic contacts. A 10nm/50nm Cr/Au layer was finally deposited as the top-gate. Final device structure is shown in Figure 1(a), which is similar to a traditional metal-oxide-semiconductor FET, but with two conducting surfaces and one conducting bulk channel in the channel region.

The crystal structure of $Bi_2Te_3$ has been shown to be similar to graphene, with stacking layers bonded by the Van der Waals force. The lattice constant for hexagonal $Bi_2Te_3$ is 0.4384 nm for the a-axis and 3.045 nm for the c-axis.[13-16] Each layer of $Bi_2Te_3$, a quintuple layer with the thickness of ~ 1nm, consists of the five layer sequence Te-Bi-Te-Bi-Te. Three quintuple layers form one unit cell.[8] On the top and bottom Te layers, there are no dangling bonds at the surface. This may lead one to suspect that it is not possible to deposit ALD $Al_2O_3$ directly on these flakes because there would be no chemical absorption due to the absence of dangling bonds at surface, as generally seen in the case of graphene. Early studies also revealed that ALD $Al_2O_3$ fails to be deposited on graphene surfaces but would only cluster and grow on graphene edges and ultimately form nanoribbons.[17-18] However, our results show that this problem does not occur for $Bi_2Te_3$. A Veeco Dimension 3100 AFM was used to study the flake surface and the layer edges after $Al_2O_3$ deposition. The pristine $Bi_2Te_3$ surface was studied after peeling and being transferred to $SiO_2$/Si substrates. We notice the surface is not as smooth as the graphene surface. This might be attributed to the fact that $Bi_2Te_3$ is not as highly air-stable as graphene; oxygen and water molecules in air can slightly oxidize the Te-terminated surface. This oxidation facilitates the following ALD process by creating nucleation spots for precursor absorption. The smallest step observed is ~ 1nm corresponding to the thickness of one quintuple layer. Most of terraces have the heights of single or multiple unit cells. Figure 1(b) shows the surface morphology of $Bi_2Te_3$ surface after 20 cycles of ALD growth using TMA and $H_2O$, corresponding to ~1.8 nm $Al_2O_3$ deposition. In the graphene case, the surface of graphene remained intact while $Al_2O_3$ nanoribbons are clearly formed at the graphene edges.[17] No $Al_2O_3$ nanoribbons or clusters can be observed at the $Bi_2Te_3$ layer edges. Figure 1(c) shows that ALD $Al_2O_3$ is conformal and uniformly coated on a series of steps of the peeled $Bi_2Te_3$ surface with the relative step height remaining similar before ALD. Figure 1(d) shows the measured heights of ~ 3nm, 9nm and 6nm corresponding to 1-3 unit cells. These observations show that the π-bond in pristine $Bi_2Te_3$ is not chemically inert in atmosphere as graphene. This instability makes direct ALD growth of dielectrics on such layered structures possible; however, the surface reactions definitely induce potential defects at the high-k/TI interface and hence deteriorate device performance as well.

[a] Author to whom correspondence should be addressed; electronic mail: yep@purdue.edu

Electrical characterization was performed after device fabrication using a Keithley 4200. Two devices, one using $TMA/H_2O$ (Device 1) and another using $TMA/O_3$ (Device 2) as precursors for $Al_2O_3$ are studied here. The typical geometric features are 2 μm in gate length and 1 μm in gate width. The channel resistance, including the contact resistance, for Devices 1 and 2 are 32kΩ and 25kΩ, respectively, indicating that the flake thicknesses are similar. Linear I-V characteristics indicate a good ohmic contact to the $Bi_2Te_3$ flakes. Figure 2(a) and (b) shows independent top-gate and back-gate modulation for both devices with another gate floating. The top-gate and back-gate leakage currents are less than $10^{-10}$A. We do not observe an electron-hole or ambipolar transition because the strong stoichiometric doping of $Bi_2Te_3$ makes it an n-type material with an estimated doping concentration of ~$10^{19}$ /$cm^3$. For Device 1, a maximum of 45.6% modulation is reached through back-gate control, where the back-gate voltage is swept from -50V to 50V. We calculate the transconductance ($g_m$) to be 9.8 nS. Comparatively, the top gate $g_m$ is measured 60 nS at its maximum, six times larger than the back-gate $g_m$ due to the thinner top dielectric thickness and higher k value. Though we get an improved value for transconductance by top-gate control, the improvement is not as high as we expect considering the oxide thickness and dielectric constant for both $SiO_2$ as back-gate oxide and $Al_2O_3$ for top-gate oxide. The top-gate $g_m$ should be around 60 times larger than back-gate $g_m$, if we don't consider top-gate partial coverage of the channel and roughly estimate the practical dielectric constant of $Al_2O_3$ to be ~7.8, twice as that of $SiO_2$. The stark contrast between predicted and measured values indicates non-ideal $Al_2O_3$/$Bi_2Te_3$ interfaces, which is consistent with the discussions above. Also, from the extrinsic transconductances, we can estimate the low limit of the effective electron mobility with top-gate control to be ~ 1.69 $cm^2$/Vs, and with back-gate control to be ~ 17.0 $cm^2$/Vs. This effective mobility contains two parts, the surface state mobility and the bulk mobility. The results indicates that the low mobility bulk channel is the dominant conducting channel and the high mobility surface channels are degraded by the poor interfaces, in particular, the top interface. Poor interface conditions easily result in a strong degradation of field-effect modulation efficiency as demonstrated in ALD high-k/III-V MOSFETs.[19] Compared to III-V MOSFETs, such interface degradation in $Bi_2Te_3$ could impose more serious problems in device performance due to its unique carrier transport properties. It has been stated in previous studies that the conductance of those topological insulators is composed of two parts: the bulk conductance and the surface conductance, including the top surface conductance and bottom surface conductance.[10] The two surface conducting channels are of interests due to its high mobility and non-scattering carrier transport. Considering the large intrinsic carrier density in its bulk, the top surface is more sensitive to the top gate control than the bottom surface, so the two surfaces do not play symmetric roles under gate bias. The bottom surface has a better interface condition with the back $SiO_2$ dielectric as it has been left intact during the fabrication process, resulting in better back-gate control at the same electrical field. However, the top surface condition of the flakes had been changed to some extent due to $Al_2O_3$ deposition, reacting with $H_2O$ at 200°C.[20] This reaction might not be severe enough to completely damage the material, as there was only a tiny trace of water vapor as one precursor in one of the alternating ALD pulses. In addition, this water corrosion could only take place in the first several cycles of $Al_2O_3$ deposition and would naturally cease when the flake is covered with $Al_2O_3$. But then again, the water corrosion could considerably impact on the interface quality, resulting in a significant decrease in surface modulation under field-effect.

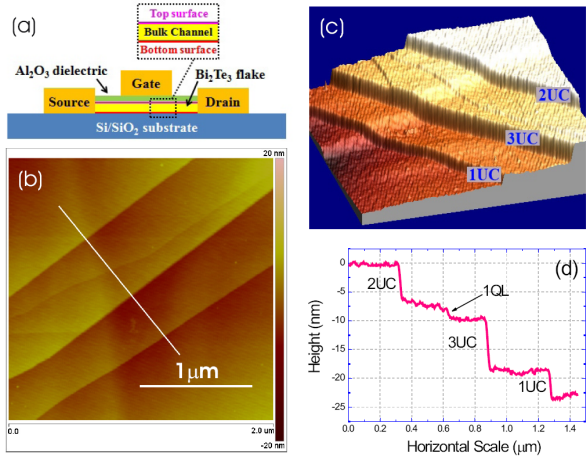

Figure 1: (a) Schematic device structure of a $Bi_2Te_3$ based TI FET. (b) AFM image of $Bi_2Te_3$ surface with multiple quintuple layer terraces coated with an ultrathin ALD $Al_2O_3$ layer; (c) Three-dimensional profile of the same surface showing different thicknesses of the broken layers with 1-3 unit cells; (d) AFM height measurement of the smooth terraces along the white line illustrated in Figure 1(b).

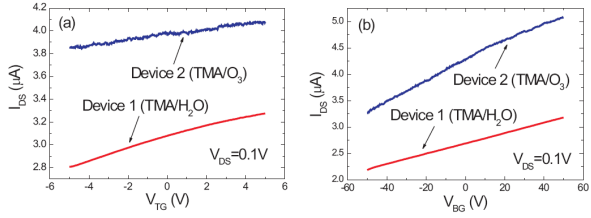

Figure 2: (a) Drain current vs top-gate bias of two $Bi_2Te_3$ TI FETs with a 10 nm $Al_2O_3$ as top-gate dielectric; (b) drain current vs back gate bias of the same devices with a 300 nm $SiO_2$ as back-gate dielectric. Drain-source voltage is 0.1V in both measurements. The observed minor drain current change at the same bias condition is ascribed to the change of contact resistance and interface traps after gate bias stress.

In the same Figure 2(a) and (b), the counterparts of top-gate and back-gate control of Device 1 are also shown. Compared to Device 1, Device 2 shows similar back-gate control, with a maximum modulation of 55.8% for a gate sweep of -50V to 50V. However, the two devices differ in top-gate control. Device 1 has a maximum modulation of 16.7% with a smoother curve whereas Device 2 has a modulation of only 5.9% with a noisier curve, with the top-gate voltage ranging from -5V to 5V. The weaker top gate modulation in Device 2 further supports our conclusion that ALD precursors are chemically absorbed to the top surface of $Bi_2Te_3$. In Device 2, the use of ozone as an oxidant ALD precursor leads to greater top surface damage during the first several cycles of ALD growth because ozone is a stronger oxidant than $H_2O$. As a consequence, the top interface of Device 2 is more defective and the top-gate shows weaker channel modulation and much noisy curves.

Finally, the simultaneous dual gate modulations of $Bi_2Te_3$ are shown in Figure 3. The top gate voltage is swept from -5V to 5V while the back gate ranges from -50V to 50V with a 20V step. We achieve the highest modulation rate of 76.1% for Device 1, and 61.8% for Device 2. All these indicate a significant enhancement in modulation for $Bi_2Te_3$ thin flakes at room temperature, using $Al_2O_3$ high-k as a top-gate dielectric and dual gate control. Development of a perfect high-k/TI interface is a must to realize real device applications based on TI FETs. In particular, the truly attractive property of TI as a channel material for device applications is its surface channel with high carrier mobility and velocity. Any formation of top-gate dielectric on semiconductors cannot be as important as on TI since the conducting channel is on the surface.

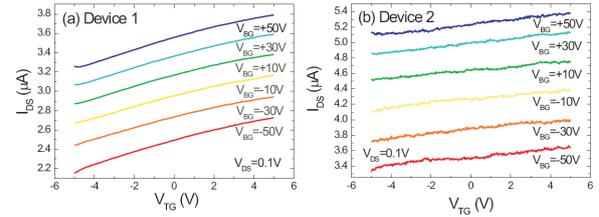

Figure 3: Drain current vs dual gate modulation of two $Bi_2Te_3$ TI FETs. (a) $Bi_2Te_3$ TI FET with a 10 nm $Al_2O_3$ deposited by TMA and $H_2O$ and (b) $Bi_2Te_3$ TI FET with 10 nm $Al_2O_3$ deposited by TMA and $O_3$.

In conclusion, we have investigated ALD high-k oxide formation on $Bi_2Te_3$ as a top-gate dielectric. AFM studies reveal the feasibility of direct ALD of high-k dielectrics on this layered material. Electrical characterization shows a pronounced modulation by both top-gate and back-gate with the highest modulation of 76.1% achieved with simultaneous dual gate control. However, at this point the top-gate modulation is not as effective as the back-gate modulation at the same electrical field due to the degraded interface between the ALD dielectric and the TI. Further studies on protecting the TI surface during ALD dielectric formation are on-going.[21-23]

The authors would like to thank X. Xu, C. Liu, G. Q. Xu, Y.P. Chen, A.T. Neal and N. Conrad for valuable discussions and E. Milligan, W. J. Qian, J. F. Tian, M. Xu for technical assistance. The work is supported by DARPA MESO program.